\documentstyle[12pt]{article}
\begin{document}
\title{The Mysterious Dark Energy}
\author{B.G. Sidharth\\
International Institute for Applicable Mathematics \& Information Sciences\\
Hyderabad (India) \& Udine (Italy)\\
B.M. Birla Science Centre, Adarsh Nagar, Hyderabad - 500 063 (India)}
\date{}
\maketitle
\begin{abstract}
The concept of an all pervading Aether is age old, and contrary to popular belief, it survived the twentieth century too though with different nuances. Using this concept of a background Quantum Vacuum, the author in 1997 proposed a cosmological model with some resemblance to the Dirac cosmology, which correctly predicted a dark energy driven accelerating universer with a small cosmological constant, as was subsequently confirmed by observation in 1998. Moreover the so called Large Number coincidences including the mysterious Weinberg formula are deduced in this theory, rather than being ad hoc. We examine the concept of Aether in this context and indicate how this dark energy may be harnessed.
\end{abstract}
\section{Introduction}
In 1997, when the ruling paradigm was dark matter and a decelarating universe, the author proposed a model based on the Zero Point Field dark energy and an accelerating Universe with a small cosmological constant. Moreover, in this theory, the empirically well known but otherwise inexplicable, so called Large Number relations, including the mysterious Weinberg formula were deduced as consequences, rather than being ad hoc \cite{x1,x2,x3,x4,x5}. In 1998, the observations of Perlmutter and others on distant supernovae confirmed the above scenario - this work was infact the Breakthrough of the Year 1998 of the American Association for Advancement of Science's Science Magazine \cite{perl,kirsh,science}. Subsequently observations by the Wilkinson Microwave Anisotropy Probe (WMAP) and the Sloan Digital Sky Survey confirmed the predominance of the new paradigmatic dark energy - this was the Breakthrough of the Year 2003 \cite{science2}.\\
We first observe that the concept of a Zero Point Field (ZPF) or Quantum Vacuum
(or Aether) is an idea whose origin can be traced back to Max Planck himself. Quantum
Field Theory attributes the ZPF to the virtual Quantum effects of an already
present electromagnetic field \cite{davydov}. What is the mysterious energy of supposedly empty vacuum?\\
It may sound contradictory to attribute energy or density to the vacuum. After all vacuum is a total void. However, over the past four hundred years, it has been realized that it may be necessary to replace the vacuum by a medium with some specific physical properties. For instance Descartes the seventeenth century French philosopher mathematician proclaimed that the so called empty space above the mercury column in a Torricelli tube, that is, what is called the Torricelli vacuum, is not a vacuum at all. Rather, he said, it was something which was neither mercury nor air, something he called aether.\\
The seventeenth century Dutch Physicist, Christian Huygens required such a non intrusive medium like aether, so that light waves could propagate through it, rather like the ripple waves on the surface of a pond. Hence the word luminiferous aether. In the nineteenth century the aether was reinvoked. Firstly in a very intuitive way Faraday could conceive of magnetic effects in vacuum in connection with his experiments on induction. Based on this, the aether was used for the propagation of electromagnetic waves in Maxwell's Theory of electromagnetism, which infact laid the stage for Special Relativity. This aether was a homogenous, invariable, non-intrusive, material medium which could be used as an absolute frame of reference atleast for certain chosen observers. However the experiments of Michelson and Morley towards the end of the nineteenth century, lead to its downfall, and thus was born Einstein's Special Theory of Relativity in which there is no such absolute frame of reference. The aether lay shattered once again.\\
Very shortly thereafter the advent of Quantum Mechanics lead to its rebirth in a new and unexpected avatar. Essentially there were two new ingredients in what is today called the Quantum Vacuum. The first was a realization that Classical Physics had allowed an assumption to slip in unnoticed: In a source or charge free "vacuum", one solution of Maxwell's Equations of electromagnetic radiation is no doubt the zero solution. But there is also a more realistic non zero solution. That is, the electromagnetic radiation does not necessarily vanish in empty space.\\
The second ingredient was the mysterious prescription of Quantum Mechanics, the Heisenberg Uncertainty Principle, according to which it would be impossible to precisely assign momentum and energy on the one hand and spacetime location on the other. Clearly the location of a vacuum with no energy or momentum cannot be specified in spacetime.\\
This leads to what is called a Zero Point Field. For instance a Harmonic Oscillator, a swinging pendulum for example, according to classical ideas has zero energy and momentum in its lowest position. But the Heisenberg Uncertainty endows it with a fluctuating energy. This fact was recognized by Einstein himself way back in 1913 who contrary to popular belief, retained the concept of aether though from a different perspective \cite{ra}. It also provides an understanding of the fluctuating electromagnetic field in vacuum.\\
From another point of view, according to classical ideas, at the absolute zero of temperature, there should not be any motion. After all the zero is when all thermodynamic motion ceases. But as Nernst, father of the third law of Thermodynamics himself noted, experimentally this is not so. There is the well known superfluidity due to Quantum Mechanical -- and not thermodynamic -- effects. This is the situation where supercooled Helium moves in a spooky fashion \cite{davydov}.\\
This mysterious Zero Point Field or Quantum Vacuum energy has since been experimentally confirmed in effects like the Casimir effect which demonstrates a force between uncharged parallel plates separated by a charge free medium, the Lamb shift which demonstrates a minute oscillation of an electron orbiting the nucleus in an atom-as if it was being buffetted by the Zero Point Field- the anomalous Quantum Mechanical gyromagnetic ratio g = 2 and so on \cite{rb}-\cite{rg},\cite{rr1}.\\
The Quantum Vacuum is a far cry however, from the passive aether of olden days. It is a violent medium in which charged particles like electrons and positrons are constantly being created and destroyed, almost instantly, infact within the limits permitted by the Heisenberg Uncertainty Principle for the violation of energy conservation. One might call the Quantum Vacuum as a new state of matter, a compromise between something and nothingness. Something which corresponds to what the Rig Veda described thousands of years ago: "Neither existence, nor non existence."\\
The Quantum Vacuum can be considered to be the lowest state of any Quantum field, having zero momentum and zero energy. The properties of the Quantum Vacuum can under certain conditions be altered, which was not the case with the erstwhile aether. In modern Particle Physics, the Quantum Vacuum is responsible for phenomena like Quark confinement, a property whereby it would be impossible to observe an independent or free Quark, the spontaneous breaking of symmetry of the electroweak theory, vacuum polarization wherein charges like electrons are surrounded by a cloud of other opposite charges tending to mask the main charge and so on. There could be regions of vacuum fluctuations comparable to the domain structures of feromagnets. In a ferromagnet, all elementary electron-magnets are aligned with their spins in a certain direction. However there could be special regions wherein the spins are aligned differently.\\
Such a Quantum Vacuum can be a source of cosmic repulsion, as pointed by Zeldovich and others \cite{rh,ri}. However a difficulty in this approach has been that the value of the cosmological constant turns out to be huge, far beyond what is observed. This has been called the cosmological constant problem \cite{rj}.\\
There is another approach,
sometimes called Stochastic Electrodynamics which treats the ZPF as primary
and attributes to it Quantum Mechanical effects \cite{r12,r13}. It may be
re-emphasized that the ZPF results in the well known experimentally verified
Casimir effect \cite{r14,r15}. We would also like to point out that
contrary to popular belief, the concept of aether has survived over the decades
through the works of Dirac, Vigier, Prigogine, String Theoriests like Wilzeck
and others \cite{r16}-\cite{r22}. As pointed out it appears that even Einstein himself
continued to believe in this concept \cite{r23}.\\
We would first like to observe
that the energy of the fluctuations in the background electromagnetic field could lead
to the formation of elementary particles. Indeed this was Einstein's belief. As Wilzeck put it, \cite{r20}, ``Einstein was not satisfied with the dualism. He wanted to regard the fields, or ethers, as primary. In his later work, he tried to find a unified field theory, in which electrons (and of course protons, and all other particles) would emerge as solutions in which energy was especially concentrated, perhaps as singularities. But his efforts in this direction did not lead to any tangible success.''\\
\section{The Zero Point Field}
 Let us see how this can happen. In the words of Wheeler \cite{rr1}, ``From the zero-point fluctuations of a single oscillator to the fluctuations of the electromagnetic field to geometrodynamic fluctuations is a natural order of progression...''\\
Let us consider, following Wheeler a harmonic oscillator in its ground state. The probability amplitude is
$$\psi (x) = \left(\frac{m\omega}{\pi \hbar}\right)^{1/4} e^{-(m\omega/2\hbar)x^2}$$
for displacement by the distance $x$ from its position of classical equilibrium. So the oscillator fluctuates over an interval
$$\Delta x \sim (\hbar/m\omega)^{1/2}$$
The electromagnetic field is an infinite collection of independent oscillators, with amplitudes $X_1,X_2$ etc. The probability for the various oscillators to have emplitudes $X_1, X_2$ and so on is the product of individual oscillator amplitudes:
$$\psi (X_1,X_2,\cdots ) = exp [-(X^2_1 + X^2_2 + \cdots)]$$
wherein there would be a suitable normalization factor. This expression gives the probability amplitude $\psi$ for a configuration $B (x,y,z)$ of the magnetic field that is described by the Fourier coefficients $X_1,X_2,\cdots$ or directly in terms of the magnetic field configuration itself we have
$$\psi (B(x,y,z)) = P exp \left(-\int \int \frac{\bf{B}(x_1)\cdot \bf{B}(x_2)}{16\pi^3\hbar cr^2_{12}} d^3x_1 d^3x_2\right).$$
$P$ being a normalization factor. Let us consider a configuration where the magnetic field is everywhere zero except in a region of dimension $l$, where it is of the order of $\sim \Delta B$. The probability amplitude for this configuration would be proportional to
$$\exp [-(\Delta B)^2 l^4/\hbar c)$$
So the energy
of fluctuation in a region of length $l$ is given by finally \cite{rr1,r24,r25}
\begin{equation}
B^2 \sim \frac{\hbar c}{l^4}\label{ea}
\end{equation}
We next argue that $l$ will be the Compton length, this is the mean length of fluctuations.  We note that as is well known, a background ZPF of the kind we have been considering can explain the Quantum Mechanical spin half as also the anomalous $g = 2$ factor for an otherwise purely classical electron \cite{sachi,boyer,taylor}. The key point here is (Cf.ref.\cite{sachi}) that the classical angular momentum $\vec r \times m \vec v$ does not satisfy the Quantum Mechanical  commutation rule for the angular momentum $\vec J$. However when we introduce the background Zero Point Field, the momentum now becomes
\begin{equation}
\vec J = \vec r \times m=\vec v + (e/2c) \vec r \times (\vec B \times \vec r) + (e/c) \vec r \times \vec A^0 ,\label{ez5}
\end{equation}
where $\vec A^0$ is the vector potential associated with the ZPF and $\vec B$ is an external magnetic field introduced merely for convenience, and which can be made vanishingly small.\\
It can be shown that $\vec J$ in (\ref{ez5}) satisfies the Quantum Mechanical commutation relation for $\vec J \times \vec J$. At the same time we can deduce from (\ref{ez5})
\begin{equation}
\langle J_z \rangle = - \frac{1}{2} \hbar \omega_0/|\omega_0|\label{ez6}
\end{equation}
Relation (\ref{ez6}) gives the correct Quantum Mechanical results referred to above.\\
From (\ref{ez5}) we can extend the arguments and also deduce that
\begin{equation}
l = \langle r^2 \rangle^{\frac{1}{2}} = \left(\frac{\hbar}{mc}\right)\label{ez7}
\end{equation}
(\ref{ez7}) shows that the mean dimension of the region in which the fluctuation contributes is of the order of the Compton wavelength of the electron. By relativistic covariance (Cf.ref.\cite{sachi}), the corresponding time scale is at the Compton scale.\\
In (1)  above if $l$ is taken to be the Compton wavelength of a typical elementary
particle, then we recover its energy $mc^2$, as can be easily verified. As mentioned Einstein himself had believed that the electron was a result
of such condensation from the background electromagnetic field (Cf.\cite{r26,ri}
for details).\\
We now very briefly indicate the cosmology referred to in the introduction (Cf.refs.\cite{x1}-\cite{x5},\cite{bgskluwer,csf}). Elementary particles are created from the ZPF as above. If there are $N$ elementary particles, then fluctuationally a nett $\sqrt{N}$ particles are created within the Compton time $\tau$ (see refs.\cite{x1}-\cite{x5}, \cite{bgskluwer} for details), so that
\begin{equation}
\frac{dN}{dt} = \frac{\sqrt{N}}{\tau}\label{eb1}
\end{equation}
We also use the well known facts that
\begin{equation}
M = Nm\label{eb2}
\end{equation}
and
\begin{equation}
R = GM/c^2\label{eb3}
\end{equation}
In (\ref{eb2}), $M$ is the mass of the Universe, $m$ the mass of a typical elementary particle like the pion, $N \sim 10^{80}$ the number of elementary particles in the Universe and $R$ its radius.\\
Differenciation of (\ref{eb3}) and use of (\ref{eb2}) and (\ref{eb1}) then leads to a host of consistent relations,
$$v = \dot{R} = HR, \quad H = \frac{c}{l} \cdot \frac{1}{\sqrt{N}},$$
$$G \rho_{vac} = \Lambda < 0(H^2), R = \sqrt{N} l, T = \sqrt{N} \tau, \rho_{vac} = \rho/\sqrt{N}$$
\begin{equation}
m = \left(\frac{H\hbar^2}{Gc}\right)^{1/3} , \frac{e^2}{Gm^2} \approx \frac{1}{\sqrt{N}}\label{eb4}
\end{equation}
and so on.\\
In (\ref{eb4}) above, $H$ is the Hubble constant, $l$ the pion Compton length, $\rho$ the average density, $\Lambda$ the cosmological constant and $\rho_{vac}$ the vacuum density. These relations include the empirically known so called Eddington formula, the Weinberg formula and the well known (but otherwise ad hoc) electromagnetism - gravitational coupling constant.\\
It may also be mentioned that all this can be interpreted elegantly in terms of underlying Planck oscillators in the Quantum Vacuum (Cf.refs.\cite{fpl2,fpl3}).\\
Finally, it may be mentioned that (\ref{eb4}) shows that both $\Lambda$ and $H \to 0$ as $N \to \infty$, as indeed is the current belief. 
\section{Harnessing the ZPF?}
Two of the earliest realisations of the ZPF as mentioned were in the form of the Lamb shift
and the Casimir effect.\\
In the case of the Lamb shift, as is well known, the motion of an orbiting
electron gets affected by the background ZPF. Effectively there is an
additioinal field, over and above that of the nucleus. This additional
potential, as is well known is given by
 \cite{bd}
$$\Delta V (\vec r) = \frac{1}{2} \langle (\Delta r)^2 \rangle \nabla^2 V
(\vec r)$$
The additional energy
$$\Delta E = \langle \Delta V (\vec r) \rangle$$
contributes to the observed Lamb shift which is $\sim 1000 mc/sec$.\\
The essential idea of the Casimir effect is that the interaction between
the ZPF and matter leads to macroscopic consequences. For example if we
consider two parallel metallic plates in a conducting box, then we should
have a Casimir force given by
 \cite{r3}
$$F = \frac{-\pi^2}{240} \frac{\hbar cA}{l^4}$$
where $A$ is the area of the plates and $l$ is the distance between them.
More generally, the Casimir force is a result of the
boundedness or deviation from a Euclidean topology in the Quantum
Vaccuum. These Casimir forces have been experimentally demonstrated
\cite{r4,r5,r6,r7}.\\
Let us return to equation (1). The ZPF fluctuations typically take place within the time $\tau$, a typical
elementary particle Compton time as suggested by (4). This begs the question whether such
ubiquotous fields could be tapped for terrestrial applications or otherwise.\\
We now invoke the well known result from macroscopic physics that the
current in a coil is given by
\begin{equation}
\imath = \frac{nBA}{r\Delta t}\label{eE}
\end{equation}
where $n$ is the numer of turns of the coil, $A$ is its area and $r$ the
resistance.\\
Introducing (\ref{ea}) into (\ref{eE}) we deduce that a coil in the ZPF
would have a fluctuating electric current given by
\begin{equation}
\imath = \frac{nA}{r} \,  \frac{e}{l^2 \tau}\label{eF}
\end{equation}
Ofcourse, this would be a small effect. But in principle it should be possible to harness the current (\ref{eF}) using advanced technologies, possibly superconducting coils to minimize $r$. 

\end{document}